\documentclass[letterpaper,twocolumn,10pt]{article}
\usepackage{usenix,epsfig,paralist,cite,url,rotating}
\usepackage[bottom]{footmisc}

\date{}

\title{\Large \bf Active~Switching:
Packet-Steering~Flow~Annotations} 

\author{{\rm Saul St.~John}\\
University of Wisconsin - Madison
\and
{\rm Aditya Akella}\\
University of Wisconsin - Madison
} 

\begin{document}
\maketitle


\subsection*{Abstract}
Our previous experience building systems for middlebox chain composition and
scaling in software-defined networks has revealed that existing mechanisms of 
flow annotation commonly do not survive middlebox-traversals, or suffer from 
extreme identifier domain limitations resulting in excessive flow table size. 
In this paper, we analyze the structural artifacts resulting in these 
challenges, and offer a framework for describing the behavior of middleboxes 
based on actions taken on traversing packets. We then present a novel mechanism
for flow annotation that features an identifier domain significantly 
larger than existing techniques, that is transparent to hosts traversed, and 
that conserves flow-table resources by requiring only a small number of 
match rules and actions in most switches. We evaluate said technique, showing
that it requires less per-switch state than conventional techniques. We then
describe extensions allowing implementation of this architecture within a broader
class of systems. Finally, we close with architectural suggestions for enabling 
straightforward integration of middleboxes within software-defined networks.


\section{Introduction}
As the enabling technology of many recent networking innovations, SDN
has been widely described and viewed as a "swiss-army knife" capable
of solving all manner of challenges within the networking domain. While
this reputation is primarily well-deserved, less discussed are the 
problem spaces in which SDN is counterproductive or of non-obvious
utility.

In previous work, we explored the challenges and opportunities of SDN
as an infrastructure for the implementation of complex traffic
intermediation and modification functionality utilizing
middleboxes~\cite{stratos}. Our experience of constructing said
controller lead us to identify a seemingly-significant, yet
little-discussed, impediment to the use of SDN for this use case as
well as for broader scenarios involving middlebox
chaining~\cite{servicechaining} and/or network functions
virtualization~\cite{nfv}.

Consider an administrative policy requiring all traffic to be shunted
through a middlebox prior to its arrival at the destination host, and
suppose there are multiple identical instances of that middlebox the
controller might choose from. In order to effect a unique per-flow
path to ensure effective load balancing, the controller must install
rules in the switches adjacent to the middleboxes and endpoints that
uniquely identify each possible traversing flow\footnote{It is
  sometimes suggested that load-balancing within an SDN can be
  achieved using prefix-match rules to decide among the candidate
  middleboxes. At best, such a solution can only provide probabilistic
  guarantees as to equitable load-distribution, and the strength of
  those guarantees are proportional to the number of prefix-match
  rules (i.e. the inverse of the cardinality of the prefix.) What's
  more, this technique is predicated on a somewhat-predictable
  distribution of source addresses within the address space. Large
  traffic spikes from a particular geographical region can easily
  break this system, saturating one device while underutilizing the
  remainder. }, leading to a state-size requirement in {\it each
  switch} that is proportional to the number of concurrent flows in
the network. This problem is usually described as {\bf state-space
  explosion}, and it presents a significant limitation to the
scalability of SDN architectures.

However, consider the possibility that the middleboxes interposed on
traffic not only inspect, but also {\it mangle} (modify) said
traffic. Such a middlebox could, conceivably, alter the traffic such
that, on egress, it no longer appears to belong to the same flow to
which it belonged on ingress (for example, by modifying header
addressing fields.) When the behavior of the middlebox is known and
deterministic, this challenge can be overcome by informing the
controller of the modifications the middlebox will make to traffic;
however, this requires middleboxes and controllers to share state, and
duplicate some work. In the case of a middlebox that might arbitrarily
and unpredictably alter traffic, though, the controller {\it cannot}
craft a match-rule that will reliably identify the flow post
middlebox-traversal. We call this problem the {\bf post-traversal flow
  reassociation problem}, and, in general, we call the challenges
involved in the implementation of such a system the {\bf
  traffic-steering problem}.

In \cite{stratos}, we partially obviated these problems by accepting
certain design and functional limitations to the behavior the controller
might effect, and by placing certain restrictions on the behavior of
middleboxes that can be utilized within the system. While functional,
the mechanisms we employed suffer from significant limitations in
scalability (e.g., imposing a fairly small limit on the number of potential
next hops in a chain) and functionality (e.g., re-purposing DSCP bits in
the IP header precludes certain forms of QoS). Similar limitations can
also be found in other approaches to this challenge (e.g., \cite{flowtags}).

This paper seeks an architecturally clean, yet immediately deployable
approach to this problem of growing importance. We present a solution
that we believe to be more flexible and more efficient than comparable
systems previously described in literature. In brief, we utilize
fields within each packet that have been rendered useless by SDN to
{\em cache} the result of the first flow-table lookup for a
packet. Put differently, we {\it annotate} each packet at the ingress
point to the network with the controller's routing decisions, thereby
eliminating the need for each packet to be matched against a
flow-unique rule at each switch in the network. We dub this technique
{\bf active switching}.

During design and implementation of our prototype active switching
controller (SOFT), we came to believe that this solution has broader
utility beyond the challenges involved in middlebox chain-composition
and traffic steering.  While the primary focus of this paper is that
set of problems, later sections describe how active switching may be
useful in different scenarios, such as the construction of network
fabrics, and routing within non-acyclic topologies.

We begin this paper by discussing the architectural artifacts that inform the 
implementation and behavior of commonly available middleboxes, and offer a 
taxonomy for describing and classifying them by their operation as it is 
visible to the network. We then discuss previously identified solutions to the 
traffic-steering problem, and their limitations. We proceed to describe the 
architecture of a novel potential solution to this problem, and evaluate a 
proof-of-concept implementation.  Finally, we close with a general sketch of 
extensions to this technique allowing it to address a broader class of problems 
to which this technique may be applicable, and take-away implications for 
various members of the networking community.


\section{Motivation}
\label{sec:motivation}
Previous work (e.g. \cite{flowtags}, \cite{stratos}) have discussed
the general set of problems related to traffic steering in detail. As such,
we do not attempt to fully describe that problem space herein, and refer the 
interested reader to related works for an in-depth treatment. Herein, we
describe the challenges only to the degree sufficient to motivate the remainder
of this paper.

\subsection{Definition}
For the purposes of this system, we define a middlebox as a network device
that interposes on network traffic prior to its receipt by the intended 
destination. This definition is somewhat surprising in its generality, in 
that switches and routers can be validly and reasonably considered to be
middleboxes under it. However, we've found that more-limited
definitions of middlebox are insufficient to describe the full spectrum
of extant devices that are traditionally considered to be one. For example,
in terms of network-visible behavior, a simple switch and a transparent
intrusion detection system operate identically, as both receive traffic
and re-emit it without modification. Similarly, a traditional IP router
and a MAC-layer network address translation device both emit traffic
received with modifications to the network addressing header fields. As
such, we believe that any reasonable definition of middlebox that
encompasses devices such as IDSs and MAC NAT must also encompass devices
such as traditional switches and routers.

\subsection{Taxonomy}
For reasons historical, there are two
predominant architectural foundations upon which the higher-level functionality
of middleboxes is implemented. In general, the distinction between these
types of middleboxes can be distinguished by the type of traffic the middlebox
is able to "see":
\begin{compactitem}
\item A \textbf{bridging middlebox} receives all packets available on the
medium.
\item A \textbf{routing middlebox} only receives packets that are constructed so
as to appear that the middlebox is a valid "next-hop" on a
route from the source address to the destination address, given the middlebox's
IP configuration.
\end{compactitem}

While this distinction implies the potential side-effects of a
given middlebox, it's important to note that it does not meaningfully describe or
constrain the behavior a given middlebox may exhibit with respect to traffic
"seen". For the purpose of discussing the behavior and higher-level 
functionality 
implemented by a middlebox, this distinction is not useful, as these types
of middlebox are roughly equipotent.

To the end of describing the behavior of middleboxes from the perspective of
the network itself, we can also classify middleboxes based on their possible
behavior \textit{upon those packets received}:
\begin{compactitem}
\item A \textbf{transparent middlebox} emits all packets received from upstream
without alteration in the downstream direction.
\item A \textbf{translucent middlebox} need not emit all packets, but all packets
that are emitted in the downstream direction are identical to
some packet received from upstream.
\item A \textbf{mangling middlebox} may emit packets downstream
that are not identical to some packet received from upstream; however, each packet 
emitted downstream "corresponds"\footnote{This relationship is left vague somewhat intentionally, as
a middlebox which appears to originate and terminate flows when considered as a black-box may, 
in fact, be recognizable as a mangling middlebox given a sufficiently 
nuanced understanding of its behavior.} to some packet received from upstream.
\item A \textbf{connection-originating middlebox} may act as an originator of
new flows.
\item A \textbf{connection-terminating middlebox} may terminate flows received.\footnote{This could also be viewed
as an extreme case of middlebox translucency.} 
\end{compactitem}

We will be using this latter terminology to describe middleboxes through which
traffic might be steered throughout the rest of the paper.\footnote{Only a
subset of this taxonomy is employed in the remainder, as only that subset
is relevant to the design of active switching. We believe though that this 
taxonomy, as a whole, could have broader utility for related middlebox 
research, and offer it to that end.}
We employ a convenient abstraction in this work, that all middleboxes have 
exactly three interfaces:
\begin{compactitem}
\item An \textbf{ingress interface}, upon which traffic is only received.
\item A \textbf{egress interface}, upon which traffic is only transmitted.
\item A \textbf{control interface}, responsible for management traffic.
\end{compactitem}
We do not believe that this abstraction in any way lessens the generality
of the discussion to follow. Consider another occasionally-useful
abstraction of interfaces, which we employ in later sections:
\begin{compactitem}
\item An \textbf{upstream interface}, upon which traffic originating from
	the external network is received, and through which traffic 
	originating from endpoint hosts within the network is transmitted.
\item A \textbf{downstream interface}, which functions as the inverse of 
	the upstream interface.
\item A \textbf{control interface}, as in the previous abstraction.
\end{compactitem}
These abstractions are equivalent, as the following construction demonstrates:
on a middlebox featuring physical upstream/downstream interfaces, the abstract
ingress interface traffic can be identified as that which is received by either
physical interface. Conversely, the abstract egress interface traffic is 
necessarily all traffic transmitted by either physical interface. The control
interface behaves identically under either abstraction.

\subsection{Topology}
\begin{figure}[h!]
\centering
\includegraphics[width=0.4\textwidth]{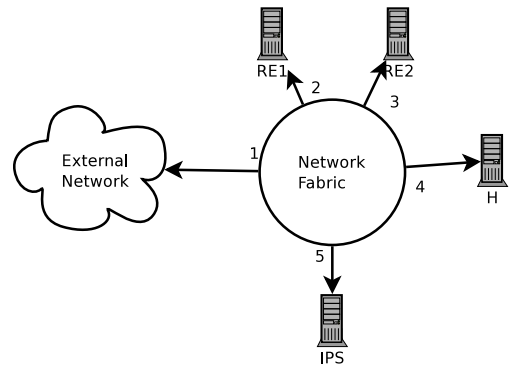}
\caption{Abstract network topology}
\label{fig:fabric}
\end{figure}
\begin{figure*}[ht!]
\centering
\includegraphics[width=0.8\textwidth]{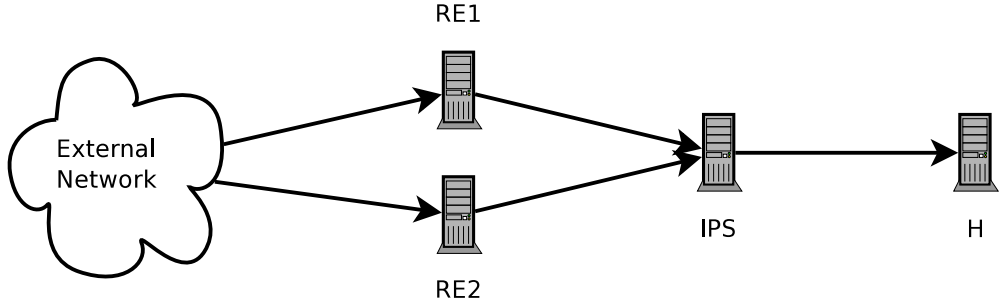}
\caption{Example policy chain}
\label{fig:policy_chain}
\end{figure*}

We assume the existence of a network fabric as described by 
M.~Casado,~\textit{et al.}in \cite{fabric}, with OpenFlow-enabled
edge switches. Positioned about the fabric is a collection of servers, 
middleboxes, and gateway switches connecting to an external network. This 
abstract topology is illustrated by Figure \ref{fig:fabric}, where the numerals
situated about the fabric identify the egress port from the network for traffic
bound for a particular device.\footnote{In later sections, we show that the
fabric abstraction is equivalent to a connected graph of OpenFlow-enabled
switches with a common controller. As such, this abstraction does not limit 
the generality or applicability of the discussion to follow.}
The figure omits, and we do not define identifiers for, the fabric's ingress
ports.

Multiple tenants may co-exist within this network; as such, the various endpoints
may not share a common owner, and their connectivity may be logically isolated
from that of other tenants within the same network by means of a network 
virtualization facility such as 802.1q~\cite{802.1q}, STT~\cite{stt}, 
VXLAN~\cite{vxlan} or NVGRE~\cite{nvgre}.

\subsection{Policy}
We further assume a policy language for describing rules interposing chains 
of middleboxes between the ingress switch and destination for traffic matching 
given patterns, and an OpenFlow controller, connected to the fabric's edge 
switches.

We will consider, as an example, a policy that specifies traffic destined for
host \textit{H} must first traverse one of two redundancy eliminators 
\textit{RE1} or \textit{RE2}, followed by a single intrusion prevention device
\textit{IPS}. This policy is illustrated in Figure \ref{fig:policy_chain}.

We further specify two general requirements on the behavior of the
system:
\begin{compactitem}
\item A flow initially submitted to one of multiple potential equivalent 
middleboxes must maintain \textbf{affinity} with that instance over the life 
of the flow.
\item A flow that passes through a given sequence of middleboxes on the 
downstream path to the destination must maintain \textbf{symmetricality} on
the return path, meaning that the very same middleboxes must be traversed in
reverse order.
\end{compactitem}

We now consider the question of how such a controller might program the edge 
switches of the fabric to effect this policy.


\section{Steering}

Previous work (e.g. \cite{flowtags}, \cite{stratos}) have discussed
the general set of problems related to traffic steering in detail. As such,
we do not attempt to fully describe that problem space herein, and refer the 
interested reader to related works for an in-depth treatment. Herein, we
describe the challenges only to the degree sufficient to motivate the remainder
of this paper.

In order for a controller to implement the desired functionality as
described in the previous section, it must be able to select, on a per-flow
basis, a sequence of intermediary devices through which traffic is shunted
prior to arrival at its destination. It must also be able to induce behavior
satisfying its policy decisions from each device along that sequence (i.e.
each network device must be able to identify the correct output port for
each flow.) Under constructions where flows are identified by a distinct 
per-switch match-rule, this rapidly leads to state-space explosion within
the network, as the number of match rules required in aggregate is proportional
to the product of the number of concurrent flows, and the number of switches
traversed by those flows.

Additionally, a controller implementing that desired functionality must also
be able to support middleboxes within the network that arbitrarily modify
traffic, without relying on an understanding of the behavior of that middlebox.
However, as such a middlebox might alter the fields of the packet upon which
match-rules rely, it does not appear that a match-rule per-switch per-flow
construction could possibly satisfy this requirement.

A number of techniques have been proposed to effect the goal of flexible
traffic steering through an SDN that do not suffer from the above-described
challenges, which we refer to generally as the {\bf traffic steering problem}. 
Some of these techniques, while superficially plausible, are insufficient for
the obviation of, or subtly still-impeded by, the traffic steering problem. 
Others impose severe limitations that constrain the utility of the approach. 

We consider these techniques and discuss their shortcomings herein.

\subsection{Policy Matching in Edge Switches}
An initially appealing, yet naive approach is to install flow rules steering 
matched traffic in each edge switch of the fabric. While this technique is 
easy to 
implement, and requires only those primitives commonly available within 
software-defined networks (e.g. those defined by the OpenFlow 1.0 standard), 
it presents a number of serious problems that effectively preclude its 
utilization in all but the most trivial of scenarios.

One such issue with this approach restricts the variety of middlebox employed
within the network, or requires undue state-sharing between all middleboxes
and the controller. We refer to this issue as the \textit{post-traversal flow
re-association} problem: during the traversal of certain
types of middlebox (specifically those that mangle L2 and/or L3 headers), it
is possible for the packet to be modified such that it no longer maintains the
expected values upon which match rules were constructed. For example, a L4-NAT
middlebox might conceivably alter all header fields upon which a traditional
OpenFlow 10-tuple might match.\footnote{This could be obviated by informing the
controller of the behavior of the middlebox in some cases, but we reject this
as a suitable general proposal for two reasons: it requires re-implementing
substantial middlebox functionality in the controller, and it doesn't work when
the behavior of a middlebox is non-deterministic, anyway.}

Supposing that the post-traversal flow annotation re-association problem were
sufficiently addressed in a given context so as to not represent a severe
limitation, another issue exists that would restrict the \textit{scalability}
of this system. In general, re-associating packets with their correct annotation
at every edge switch requires each switch to maintain at least as many flow-
table entries as there are flows that might traverse the switch at any given 
time.\footnote{In fact, the most straightforward implementation requires even 
more state than that: each switch would need an ingress rule and an egress rule
for traffic entering and departing the middlebox, respectively.} While there 
are specific situations where a sufficiently informed controller
could craft flow-table entries that correctly match on more than one flow,
such a technique is not generally applicable, and still suffers from the same
flaw: the amount of state in each switch remains proportional to the number
of flows that might be expected to traverse it, and, therefore, the number
of flows supported across any given switch along the edge of the fabric is 
limited by the size of the flow table of that switch.

\subsection{Packet Tagging in Ingress Switches}

We now consider the viability of techniques that require state proportional
to the number of current flows at ingress switches only.\footnote{We believe
this to be a more tractable scalability limitation, as multiple ingress
switches could be employed simultaneously in parallel, or switches with larger
flow tables could be used, without requiring such capability at \textit{every}
edge-switch in the network.}

\subsubsection*{Tunneling}
"Tunneling" protocols such as VXLAN~\cite{802.1q}, STT~\cite{stt}, or 
NVGRE~\cite{nvgre} are often proposed (in a hand-wavey fashion) as potential 
tools to steer traffic among middleboxes. Upon more serious consideration, 
however, a number of flaws to these approaches become apparent.

When considering the case of a tenant with a single, "linear" policy chain,
these technologies may appear to offer a plausible solution to the 
traffic steering issue. However, when considering the case of a policy-chain
with multiple potential next-hops at any point along the path, it is obvious 
that this approach suffers
from the same post-traversal flow annotation re-association problem discussed
previously.\footnote{Consider also the case of a tenant with multiple linear
policy chains that share a subset of middleboxes, or that describe policy
requiring traversal in a different order.}

Fundamentally, these techniques operate by assigning a unique identifier to
each tunnel or virtualized network existing upon the underlying network, that
is carried by all packets belonging to these tunnels. As the
identifiers serve to allow the underlying network to differentiate between 
individual links or tenants, it does not seem plausible that the very same 
identifier could also be reasonably extended to differentiate between paths
\textit{within} the overlay.

We believe that tunneling protocols are not appropriate tools to this
end, as their utilization for this purpose conflates the goals of tenant 
isolation and traffic steering. Their design was motivated by the need to 
provide the appearance of an isolated broadcast domain to tenants. Given that
the challenges around traffic-steering are not at all mitigated by the
existence of an isolated broadcast domain, it does not seem plausible that 
tools designed to effect an isolated broadcast domain would be sufficient to 
meet this need.

\subsubsection*{DiffServ Code Point Repurposing}
\label{sec:global_dscp}
Prior work (e.g. \cite{flowtags}, \cite{stratos}) has proposed repurposing
the DSCP bits in the IP header to the end of annotating packets for correct
steering. While this approach can work in practice, it, too, suffers from many
of the limitations previously described: for example, the number of flows such 
a technique is capable of supporting concurrently is constrained by 
the number of annotations that might be encoded with the 6 DSCP bits 
($2^6=64$).

More significantly, the correctness of this approach is predicated on a pair 
of brittle and potentially unsafe assumptions: that the DSCP bits are otherwise
unused within the network, and that middleboxes emit them unmolested. The 
former assumption precludes the use of QoS within the network, and the latter
may simply not be true. 

\subsection{Other techniques}
Related work has suggested the use of shims between the L2 and L3 header, or 
requiring modifications to middleboxes that
are to be employed within the SDN. We reject the former approach for suffering
from the same limitations as tunneling-based solutions, and the latter for
not being generally applicable.

The elimination of middleboxes entirely has also been suggested, by, e.g., 
\cite{end_to_middle}. To an extent, we \textit{agree} that this is
a valid approach in some cases: software-defined networking has made 
available, \textit{within
the network}, sufficient functionality to effect many network-layer tasks
traditionally performed by middleboxes (e.g., NAT). We believe that the
responsibility for such functionality should be devolved to the network itself
where possible, and that such devolution is entirely 
appropriate. However, there exist many tasks for which middleboxes are commonly
used that do not lend themselves to being integrated into the network proper,
as they require actions to be taken based on an understanding of upper-layer
protocols. It does not seem plausible to assume that all middlebox functionality
possible could ever be devolved to the network entirely, and, as such, we 
dismiss this approach as fantasy.\footnote{We have, however, limited the
types of middleboxes under consideration in this paper to those which cannot
be adequately handled by the network itself, as a result of this argument.}

\subsection{Active Switching}

The remainder of this paper discusses an approach to traffic steering that
we believe to be novel as a whole, although inspired by various aspects of 
the techniques previously discussed. We call this method \textit{active
switching}, as it requires that edge switches to act on packets in ways
other than just forwarding based on match-rule logic. We believe that it
substantially mitigates the flaws of the previously discussed possibilities.


\section{Construction}
\label{sec:const_basic}

Active switching's design is inspired by the following observations:
\begin{compactitem}
\item OpenFlow-enabled switches need not be compliant with IEEE 802.1d, and,
	by default, are not. In other words, "this ain't DIX Ethernet."
\item As OpenFlow-enabled switches are able to forward traffic based on
	L3 header match, L2 addressing need not uniquely identify a network
	host.
\item In fact, L2 addressing conveys no additional information to the network
	that could not be gleaned by examining L3 headers and network topology.
\end{compactitem}

Given those observations, we have chosen to re-purpose the source and destination
MAC address fields in the Ethernet frame header to encode the policy-defined
hop-by-hop path a packet should follow between the ingress gateway and the
destination endpoint host.

\subsection{Required flow-rule actions}
We require three register extensions to the OpenFlow 1.0 standard: 
\begin{compactitem}
	\item bit-oriented partial load from field offset to register
	\item bit-oriented partial save to field offset from register
	\item output to port given in register
\end{compactitem}
Given that fabric-edge switches in cloud architectures tend to be implemented
in software, and as these actions are already supported by Open vSwitch 
\cite{ovs}, we do not believe that these requirements impose a significant 
limitation to this 
approach. \footnote{There may also be designs employing features of
the OpenFlow 1.3 standard that can achieve reasonable facsimiles of this 
functionality, possibly at the cost of increased flow-table size.}

\subsection{Baseline Functionality}
We initially consider only the composition of middleboxes that do not mangle
L2 traffic headers, and that do not originate connections. We additionally
restrict the port identifiers of the fabric to 8 bits, and exclude paths
with a hop-count greater than $5$ from consideration.

In section \ref{sec:ext}, we will loosen these restrictions in order 
to better support a greater variety of middleboxes under less-abstract network 
topologies.

\subsubsection{Ingress Switch Behavior}
\label{sec:ingress_behavior}
The default action for packets received by an ingress switch that fail to match
any current flow rules is to forward the packet to the controller. 

Upon receipt of a packet from an ingress switch, the controller considers its 
configuration, policy, and knowledge of the network topology to construct a 
port-by-port path through the network. It then writes a flow-rule into the 
ingress switch that matches the flow under consideration and causes the
destination MAC address to be rewritten. We encode the hop-by-hop path 
through the network as follows:

\newcommand{\minisection}[1]{\smallskip \noindent{\bf #1.}}
\newcounter{linenum}
\newenvironment{proto}{\setcounter{linenum}{0}\begin{tabbing}\makebox[-0.02in]{}\=\+\kill}{\end{tabbing}}
\newenvironment{algorithm}[2]{\setcounter{linenum}{0}\begin{tabbing}\textsc{#1}\((#2)\)\\
\makebox[0.2in]{}\=\+\kill}{\end{tabbing}}
\newcommand{\all}[1]{\addtocounter{linenum}{1}\'#1\\}

\begin{figure}[h]
\centerline{
\setlength{\tabcolsep}{0pt}
\framebox[1\columnwidth][l]{
\vspace{1em}
\begin{minipage}[t]{1\columnwidth}
{\small
\begin{proto}
xxx\=xx\=xx\=xx\=xx\=xx\=xx\=xx\=xx\=xx\=xx\=xx\= \kill
\all{{\bf dst\_construct($\mathit{Path}\ P$)}}
\all{1\> $\mathit{dst\_mac} \leftarrow \mathit{00:00:00:00:00:FF}$}
\all{2\> for $\mathit{h} \in $reverse$(P)$:}
\all{3\>\> $\mathit{dst\_mac} \leftarrow ($octet\_left\_shift$(
	\mathit{dst\_mac}) \mathrel{\&} \mathit{h})$}
\end{proto}
}
\end{minipage}
\vspace{-3em}
}
}
\caption{Destination MAC address construction}
\vspace{-1em}
\label{algo:dst_constr}
\end{figure}

Subsequent to destination MAC address re-writing, the packet is handled as
though it were received by an edge switch.

\subsubsection{Edge Switch Behavior}
\label{sec:edge_behavior}
Given the register extensions previously discussed, the operation of edge-
switches is quite simple: upon receipt of a packet, the switch rewrites
the destination MAC address by shifting it one octet right. The packet is
then output through the port identified by the byte that was shifted off
the destination MAC address.

\begin{figure}[h]
\centerline{
\setlength{\tabcolsep}{0pt}
\framebox[1\columnwidth][l]{
\vspace{1em}
\begin{minipage}[t]{1\columnwidth}
{\small
\begin{proto}
xxx\=xx\=xx\=xx\=xx\=xx\=xx\=xx\=xx\=xx\=xx\=xx\= \kill
\all{{\bf octet\_rshift\_field$(\mathit{Field}\ F$)}}
\all{1\> let $\mathit{R} := $allocate\_and\_zero\_register$()$}
\all{2\> $\mathit{R} \leftarrow $load\_field$(F, 8, $len$(F) - 8)$}
\all{3\> $F \leftarrow \mathit{R}$}
\all{}
\all{{\bf handle\_packet$(\mathit{Packet}\ P)$}}
\all{4\> let $\mathit{F} := $P$.$dst\_mac}
\all{5\> let $\mathit{R} := $allocate\_and\_zero\_register$()$}
\all{6\> $\mathit{R} \leftarrow $load\_field$(F, 0, 8)$}
\all{7\> octet\_rshift\_field$(F)$}
\all{8\> output\_to\_port$(P, R)$}
\end{proto}
}   
\end{minipage}
\vspace{-3em}
}
}
\caption{Basic edge-switch logic}
\vspace{-1em}
\label{algo:handle_packet}
\end{figure}

\subsubsection{Upstream path}

While constructing the downstream path through the network, the controller
can also construct a reverse path from the endpoint back to the ingress 
switch by simply reversing the hop order and pushing a flow-rule into
the switch adjacent to the endpoint. Alternatively, the controller could
become involved in processing the first packet of a flow in either direction,
in which case similar logic would be employed, with the conceptual ingress-
point for the unidirectional flow being the switch adjacent to the end host.

In either case, the only additional logic required is a rule at the gateway
to the external network rewriting all destination MAC addresses to that of
the upstream external router.\footnote{Assuming, of course, that the link
between the gateway and the upstream router is, in fact, traditional Ethernet.}

\subsubsection{Example}
Consider the topology and policy described in section \ref{sec:motivation}.

Upon receipt of the first packet of a new flow from the external network that
is addressed for receipt by host H, the ingress switch discovers that the 
packet received fails to match any existing flow-rules, and therefore sends it 
to the controller. Using its knowledge of the network's topology and the 
policy configuration, the controller selects a redundancy eliminator for 
handling the flow. Suppose it selected RE1.

The controller will then construct the destination MAC address 
00:00:FF:04:05:02, and program the ingress switch to rewrite packets for this
flow accordingly. From this point forward, the packet is processed by edge
switch logic only.

The initial traversal of the fabric results in the destination MAC address
being re-written to 00:00:00:FF:04:05, and that altered packet to be output
from port 2 (to which RE1 is connected.) As the middlebox does not mangle L2
headers, that address will remain intact when the packet is re-received by
the fabric. It will then be output from port 5, to the IPS, and, finally,
from port 4, the endpoint for which the packet was originally destined.
\footnote{In order for H to \textit{receive} this packet, the destination
MAC address on the packet must be identical to the hardware address of H's
receiving interface. This is trivial, as modern interfaces near-universally 
support administrator-configured Ethernet addresses. We assume herein that all 
interfaces connected to an actively switched network will be configured
with the Ethernet address "00:00:00:00:00:FF", and all ARP queries will receive
a response indicating that address.\footnotemark}
\footnotetext{\label{foot:arp_abuse} One alternative we have explored
is to have the controller respond to ARP requests directly with a path-encoded
MAC address, rather than rewriting flows in the edge-switch. This mechanism can 
be useful, in that it reduces controller overhead substantially. It comes, 
however, at the cost of flexibility: individual flows cannot be independently
steered; all traffic between two endpoints must follow the same path, as ARP
requests are only issued on a per-host basis. What's more, the controller may 
not be able to invalidate a network participant's cached ARP-table entry:
although it seems reasonable that an unsolicited ARP reply should suffice, we've
observed that default security policy on many devices cause such packets to be
ignored.}

Subsequent packets of the flow received at the ingress switch will be annotated
and steered identically, without controller involvement.


\section{Supporting arbitrary middleboxes}
\label{sec:ext}
In previous sections, we limited those middleboxes supported within an active
switching architecture to only those that do not modify the network address
fields of traversing packets, and maintain a one-to-one correspondence between
packets received and packets emitted. Those limitations, while helpful in describing 
the design of this system, dramatically restrict the variety of middleboxes that
can be supported.

Herein we present extensions to the logic described by previous sections that
can be employed to the end of supporting arbitrary middleboxes, {\it including}
those that mangle MAC addresses and originate flows.

\subsection{Flow-originating middleboxes}
Middleboxes that originate new flows can be trivially supported within this
architecture via logic similar to that employed for reverse-path construction.
Conceptually, each edge-switch adjacent to a middlebox from whence a flow 
might originate is treated as an ingress switch: the controller produces a
path annotation for the initial packet of the flow, and subsequent packets
are annotated directly in the edge switch. Each edge switch so used must then
maintain state of size $O(n)$, where $n$ is again the number of flows 
originating at the device adjacent to the switch.

\subsection{L2-mangling middleboxes}
In order to support middleboxes that are not L2-transparent, it's useful to 
take a step back and consider the purpose of L2-header mangling. There are two 
cases to consider:

\subsubsection{Middleboxes providing network-layer service}
\label{sec:routers_with_side_effect}
By "network-layer service," we are referring to functionality that acts on
packets' layer 2 headers exclusively. This would be applications such
as ARP-spoofing detection, or ether-NAT.

Active switching cannot support integration of such middleboxes within
the architecture. In many cases, the functionality provided by such middleboxes
is not applicable to networks other than traditional Ethernet. As
software-defined networking provides requisite primitives to effect the 
functionality of the remainder directly within the network, we do not believe 
that the lack of support for this class of 
middlebox represents a significant limitation of the architecture.

\subsubsection{Middleboxes providing application-layer service}
\label{sec:middleboxbox}
We describe a middlebox that coerces L2-transparency from middleboxes 
implemented as routers.\footnote{Note that, while the description herein
refers to the encapsulation layer as a "middlebox", it need not be a discrete
network participant. In the proof-of-concept implementation described in
section \ref{sec:poc}, this middlebox is implemented as a dedicated OpenFlow
switch; it would be feasible furthermore to fully integrate this functionality
directly into the network edge switches.} 

\begin{figure}[b]
\centering
\includegraphics[width=0.4\textwidth]{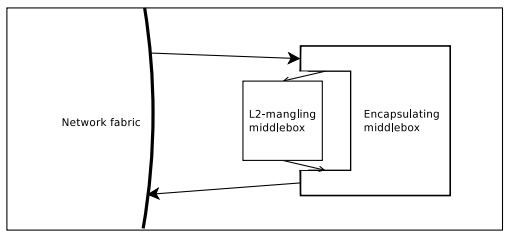}
\caption{Encapsulating middlebox connectivity}
\label{fig:middleboxbox}
\end{figure}

This middlebox has four logical interfaces:
\begin{compactitem}
	\item upstream outer
	\item upstream inner
	\item downstream outer
	\item downstream inner
\end{compactitem}

The outer interfaces connect to the edge of the switching fabric in the 
directions indicated. The inner interfaces connect to the encapsulated
middlebox's interfaces in the directions indicated. These connections
are illustrated in Figure \ref{fig:middleboxbox}.

There are several plausible mechanisms by which this encapsulation layer
can address the post-traversal flow re-annotation challenge resulting from
the encapsulated middlebox's L2-header mangling. We present two alternatives:
the first requires some amount of semantic understanding of the behavior of
the middlebox; the second does not, but constrains the number of possible
path continuations of flows traversing the middlebox.

\paragraph{Associative array.} %
Upon receipt of a packet on the upstream outer interface, the encapsulating 
middlebox extracts sufficient information from the packet to identify it when
emitted from the encapsulated middlebox in the downstream direction, and uses 
that information as a key into an associative array storing the packet's L2
addressing. The encapsulating middlebox then mangles the L2 headers as/if 
required by the middlebox encapsulated, and emits that packet on the upstream 
inner interface.

Upon receipt of a packet on the downstream inner interface, the
middlebox extracts from the packet the information required to
dereference the associative array, and rewrites the packet with the L2
headers thereby produced before emitting it on the downstream outer interface.

\paragraph{Local DSCP tagging.} \label{sec:local_dscp} %
This mechanism operates similarly to the previous, with one major exception:
rather than extracting information from the packet to key the array, we instead
assign an identifying tag to each path continuation observed on packets prior
to traversing the interior middlebox, and tag the packet with said identifier
using the DSCP bits.

Using the DSCP bits in this fashion substantially mitigates the drawbacks
described in section \ref{sec:global_dscp}, as tags need not be globally
unique, nor have consistent meaning at different points in the network. In
fact, this technique can be employed even when the network as a whole
utilizes the DSCP bits for QoS, although it further constrains the number 
of possible path continuations beyond the encapsulation layer.


\section{Extensions}

We present a number of example extensions to the active switching 
architecture that might be employed to support its use in service of 
policies requiring longer path-lengths, and upon network topologies other 
than an abstract fabric.

\subsection{Path length}

There are a number of mechanisms that can be employed to effect traffic-
steering over paths longer than five hops.

\subsubsection{Address swapping}
In order to support paths of up to 10 hops, simple changes are required
to the algorithms presented in sections \ref{sec:ingress_behavior} and
\ref{sec:edge_behavior}. 

In order to encode hops six through ten, we can also utilize the source
address field in the Ethernet header. We omit a detailed description of
the construction algorithm, as it is obvious.

We present the modified edge-switch behavior in figure 
\ref{algo:handle_packet_x}.
\begin{figure}[h]
\centerline{
\setlength{\tabcolsep}{0pt}
\framebox[1\columnwidth][l]{
\vspace{1em}
\begin{minipage}[t]{1\columnwidth}
{\small
\begin{proto}
xxx\=xx\=xx\=xx\=xx\=xx\=xx\=xx\=xx\=xx\=xx\=xx\= \kill
\all{{\bf handle\_packet$(\mathit{Packet}\ P)$}}
\all{1\> let $\mathit{F} := $P$.$dst\_mac}
\all{2\> let $\mathit{R} := $allocate\_and\_zero\_register$()$}
\all{3\> $\mathit{R} \leftarrow $load\_field$(F, 0, 8)$}
\all{4\> if $R == 0xFE$:}
\all{5\>\> let $S := $P$.$src\_mac}
\all{6\>\> $R \leftarrow $load\_field$(S, 0, 48)$}
\all{7\>\> $F \leftarrow \mathit{R}$}
\all{8\>\> $S \leftarrow 0$}
\all{9\>\> handle\_packet$(P)$}
\all{10\> else:}
\all{11\>\> octet\_rshift\_field$(F)$}
\all{12\>\> output\_to\_port$(P, R)$}
\end{proto}
}   
\end{minipage}
\vspace{-3em}
}
}
\caption{Edge-switch logic supporting 10-hop paths}
\vspace{-1em}
\label{algo:handle_packet_x}
\end{figure}

\subsubsection{Flow re-annotation}

To handle circumstances where even a 10 hop path is insufficient to effect
steering policy, a number of techniques may be employed to re-annotate
flows within the network.
\footnote{These examples are \textbf{not suggestions};
they are provided only to illustrate the potential of active switching. The 
author respectfully submits that traffic steering over paths longer than
10 hops is a solution in search of a problem.}
Three options are presented herein; the first fails
to meet the goal of supporting arbitrarily long paths, while the latter two 
represent the extrema of a spectrum of possibilities trading switch state 
and performance.

\paragraph{Alternative storage.} Conceivably, a number of other writable header
fields within the packet could be appropriated to encode path extensions.
This method would require a deeper understanding of the behavior of the
middleboxes along the path, however, and would still impose a constant 
upper-bound on the length of a path that a packet might be steered through; 
we thus reject it as an insufficient solution.

\paragraph{Table lookup.} When the number of possible path continuations from
an edge switch is 
small, the final octet of a MAC address can be used as an index into a lookup-
table of possibilities. This requires that the controller's behavior upon
receipt of a packet be modified such that, if a packet will require in-network
re-annotation, an identifier $i$ is allocated; and a flow rule installed at
the switch adjacent to the 10th hop re-annotating packets matching 
00:00:00:00:00:$i$.

When the number of path continuations is large, match rules against L3 headers
may be employed to re-annotate packets.\footnote{We suggest it plausible that
a combination approach, where $i$ indicates that a packet should be
re-submitted through a particular flow table containing L3-match rules, might
yield a performance enhancement.}

\paragraph{Controller involved.} Trivially, every edge switch could submit
all packets with empty source and destination addresses to the controller
for reconsideration. This might be necessary if, for example, the path
continuation of a flow beyond a certain point can not be ascertained by the 
controller prior to the packet's processing by some prior middlebox.

\subsection{Larger Fabrics}

The only challenge with supporting fabrics exposing more than 255 ports is
that we cannot conveniently encode port identifiers in a single octet. It is,
however, straightforward to modify the techniques described previously to 
consider, e.g., pairs of bytes to identify a port, at the cost of path-length, 
when the number of ports on the fabric is known or bounded.
\footnote{Conceivably, an encoding could be developed that did not assume
fixed-width port identifiers as well, although the edge-switch logic
to support such a scheme appears daunting.}
The details are omitted for brevity.

\subsection{Alternate topologies}
The abstraction of a network fabric is convenient for these purposes, because
it allows us to cleanly differentiate between inter-hop and intra-hop routing.
However, active switching can be used to effect intra-hop traffic-steering
as well.

We assume a network topology consisting of an 
ingress switch, and no more than fifteen interior 32-port switches, 
connected so as to form 
a maximally-interconnected graph. To each interior switch are connected no more
than fifteen middlebox or endpoint devices. \footnote{It's interesting to note 
that a fully-connected mesh of
switches where there is only \textbf{one} middlebox or endpoint
device connected to each interior switch is indistinguishable from
a fabric for the purposes of this construction.}
The ingress switch connects to an external network, as before.
\begin{figure}
\centering
\includegraphics[width=0.4\textwidth]{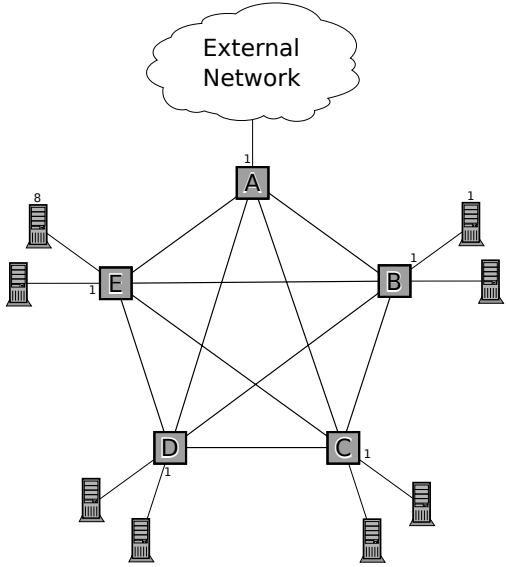}
\caption{Example mesh network topology}
\label{fig:mesh_topo}
\end{figure}
An example of such a topology is shown in figure \ref{fig:mesh_topo}. On each
switch, only the port with identifier 1 is shown; the identifiers of subsequent
ports are sequential, moving in the clockwise direction.

We present two techniques for traffic steering through such a topology.

\subsubsection{Hop-by-hop}
\label{sec:hop_by_hop}
Under the given topology, all traffic at any point in the network is
no more than two switch traversals away from its next hop. What's more, all
traffic received by any switch from an endpoint device or middlebox 
must traverse at most one switch, at which point one of the attached
devices must be the next hop. We can exploit these known constraints to 
encode a five-hop path into a MAC address by using nibbles to identify
the egress port from whence a packet should be transmitted. As there
are no more than sixteen possible next-hop switches, and each switch connects
to no more than 15 endpoint devices, this representation is non-ambiguous.

\paragraph{Example.} \label{p:hop_by_hop_ex} %
Suppose policy specified that traffic received at the
ingress switch and addressed for receipt by host 8 should be first steered
through host 1. The first packet of such a flow would be sent to the controller,
which would determine that the sequence of ports through which the flow should
be emitted are as follows: from switch A's port 2, then from switch B's port 1
(at which point the flow will traverse host 1), then from switch B's port 5,
and finally from switch E's port 2 (to host 8). The controller, using this
information, constructs the flow annotation "00:00:00:FF:25:12", and installs 
a flow-rule tagging this flow with that destination address. At each switch
in the network, the least significant nibble is shifted off the address, and the
packet is emitted out the port so identified.

\subsubsection{Destination encoding}
When dealing with fabrics, we claimed to be encoding port identifiers into the 
MAC address of packets. In order to extend that port-by-port traffic steering 
concept to non-fabric topologies, in the previous section we encoded the egress
port from each switch into the MAC address.

However, implicit in the fabric abstraction was a one-to-one correspondence
between port identifiers and the devices situated adjacent to those ports. It
is as reasonable to claim that the fabric techniques encoded not port 
identifiers, but device identifiers, into the packet itself. We can extend this
concept to non-fabric networks when the set of endpoint devices and
middleboxes in the network is fixed, and the topology is known. The trade-off
is that this requires maintaining $O(n)$-size state in each switch, where
$n$ is the number of devices in the network.

We begin by assigning a unique identifier to the ingress switch, and to each
middlebox and endpoint device. As in section \ref{sec:const_basic}, the 
controller will construct MAC addresses encoding the identifiers of each 
middlebox through which traffic will be steered prior to arrival at the flow 
endpoint.

In order to preserve identifiers over, potentially, multiple inter-switch
links, the flow-table actions must be altered. In figure 
\ref{algo:mesh_prog}, we present an algorithm for the controller to program
the switches in the network, assuming the existence of functionality to
interrogate the network topology, and global arrays of network device 
identifiers and switches. 

The function "install\_rewrite\_rule" installs a flow-table rule that operates
as described in section \ref{sec:edge_behavior}. The function 
"install\_forward\_rule" installs a rule matching the final octet of 
the destination
MAC address against the given identifier, and forwards the packet out the
given port unmodified. 

Effectively, we build a forwarding table for all identifiers in the network,
for each switch in the network.

\begin{figure}[h]
\centerline{
\setlength{\tabcolsep}{0pt}
\framebox[1\columnwidth][l]{
\vspace{1em}
\begin{minipage}[t]{1\columnwidth}
{\small
\begin{proto}
xxx\=xx\=xx\=xx\=xx\=xx\=xx\=xx\=xx\=xx\=xx\=xx\= \kill
\all{{\bf program\_switch$(\mathit{Switch}\ S)$}}
\all{1\> for $\mathit{id} \in \mathit{ALL\_IDS}$:}
\all{2\>\> if is\_adjacent$(S, \mathit{id})$:}
\all{3\>\>\> install\_rewrite\_rule$(S, \mathit{id}, $get\_port\_for\_id$(
S, \mathit{id}))$}
\all{4\>\> else:}
\all{5\>\>\> install\_forward\_rule$(S, \mathit{id}, $get\_next\_hop$(
S, \mathit{id}))$}
\all{}
\all{{\bf program\_network$()$}}
\all{6\> for $\mathit{s} \in \mathit{ALL\_SWITCHES}$:}
\all{7\>\> program\_switch$(\mathit{s})$}
\end{proto}
}   
\end{minipage}
\vspace{-3em}
}
}
\caption{Controller algorithm for mesh-topology programming}
\vspace{-1em}
\label{algo:mesh_prog}
\end{figure}

\paragraph{Example.} Suppose the policy described in the example from section 
\ref{p:hop_by_hop_ex}. Suppose further that the next-hop tables
shown in table \ref{tab:next_hop_table} have been already installed by the 
controller in each switch, such that destination MAC address is only shifted
when a switch outputs a packet from ports 1 or 2.

The controller simply constructs the MAC address 00:00:00:FF:08:01. Switch A
outputs the packet unmodified on port 2, leading to switch B, which shifts
the MAC address right by an octet, and outputs the packet on port 1. Upon
re-receipt by switch B, it is emitted on port 5 towards switch E, unaltered.
Switch E shifts the MAC address right by an octet, and emits the packet on port
2.

\begin{table}
\begin{center}
\textbf{Destination ID}
\begin{tabular}{l c || c | c | c | c | c | c | c | c | c}
	\cline{2-11}
	&  & gw & 1 & 2 & 3 & 4 & 5 & 6 & 7 & 8 \\
	\cline{2-11} \cline{2-11}
	& A & 1 & 2 & 2 & 3 & 3 & 4 & 4 & 5 & 5 \\
	\cline{2-11}
	& B & 6 & 1 & 2 & 3 & 3 & 4 & 4 & 5 & 5 \\
	\cline{2-11}
	& C & 5 & 6 & 6 & 1 & 2 & 3 & 3 & 4 & 4 \\
	\cline{2-11}
	& D & 4 & 5 & 5 & 6 & 6 & 1 & 2 & 3 & 3 \\
	\cline{2-11}
	\begin{rotate}{90}\textbf{Source~Switch}\end{rotate} & %
	E & 3 & 4 & 4 & 5 & 5 & 6 & 6 & 1 & 2 \\
	\cline{2-11}
\end{tabular}
\end{center}
\label{tab:next_hop_table}
\caption{Next hop table showing per-switch egress port for each destination ID.}
\end{table}


\section{Implementation}
\label{sec:poc}
We have implemented the functionality described in section 
\ref{sec:hop_by_hop} as a proof-of-concept, subject to the alteration described
in footnote \ref{foot:arp_abuse}. The implementation is written in Python, and 
contains approximately 500 lines of code. It is constructed as a module for the 
POX~\cite{pox} controller, and has been successfully deployed on a 
Mininet~\cite{mininet} testbed utilizing Open vSwitch.

We have additionally implemented the encapsulation layer described in section
\ref{sec:local_dscp}, so as to support middleboxes that act as routers.



\section{Analysis}

We consider an abstract network fabric and network controller to demonstrate
the state-space advantage resulting from active switching compared to
traditional match-and-forward logic. To the network fabric are attached six
servers and an external connection. Of those six servers, two are
connection-terminating endpoints (e.g. web servers), and the remaining four
consist of two pairs of equivalent middleboxes. The controller's configured
policy is such that each new flow received from the external connection is
randomly assigned to one of each type of middlebox, and one endpoint server.

Under the traditional paradigm, each new flow would result in a number of 
new rules being installed in all switches along the flow's path: a match rule
forwarding ingress traffic to the correct first middlebox's edge switch,
no less than four rules in each middlebox-attached edge switch (received from
upstream, middlebox-bound; received from middlebox, downstream-bound; etc),
and two rules in the endpoint device-attached switch. As each rule would need
to uniquely identify a flow, multiple flows cannot be handled by the same rule;
every new flow results in at least eleven new flow-rules being installed in the
network, and control messages being sent to at least four switches.

\begin{table}[h!]
	\begin{center}
		\center{\textbf{Match-and-Forward}} \\
		\begin{tabular}{|c || c | c | c | c|}
			\hline
			concurrent flows & 1 & 10 & 100 & n\\
			\hline
			ingress rules & 2 & 11 & 101 & $ n + 1 $ \\
			middlebox rules & 8 & 80 & 800 & $ 8n $ \\
			endpoint rules & 2 & 20 & 200 & $ 2n $ \\
			\hline
			total & 12 & 111 & 1101 & $ 11n + 1 $ \\
			\hline
		\end{tabular}
	\end{center}
	\begin{center}
		\center{\textbf{Active Switching}} \\
		\begin{tabular}{|c || c | c | c | c|}
			\hline
			concurrent flows & 1 & 10 & 100 & n\\
			\hline
			ingress rules & 2 & 11 & 101 & $n + 1$ \\
			middlebox rules & 1 & 1 & 1 & $1$ \\
			endpoint rules & 1 & 10 & 100 & $n$ \\
			\hline
			total & 4 & 22 & 202 & $ 2n + 2 $ \\
			\hline
		\end{tabular}
	\end{center}
	\label{tab:space_comparison}
	\caption{Comparison of flow-table space requirements using traditional and active switching logic.}
\end{table}

Using active switching, we can support an arbitrary number of upto 5-hop
paths through a fabric with upto 254 ports. Each edge switch adjacent to
an endpoint, and the ingress switch, must maintain state of size $O(n)$,
where $n$ is the number of flows that originate from the adjacent device.
All other edge-switches, such as those adjacent to middleboxes, 
need only maintain constant-size ($O(1)$) state.

\begin{figure}[h!]
\begin{center}
	\input{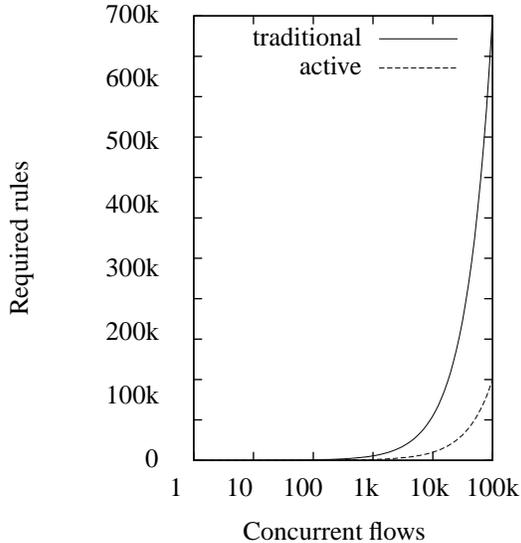}
\end{center}
\label{fig:space_comparison}
\caption{Flow-table entries required as a function of concurrent demand.}
\end{figure}

To the best of our knowledge, no existing techniques can make such claims.


\section{Implications}

We arrived at a number of unexpected implications relevant to various subsets
of the networking community in the course of this research. We offer them 
herein.

\paragraph{For middlebox designers.} Perhaps the most frustrating aspect of this
work has been attempting to integrate L2-mangling middleboxes (or, as we have
come to describe them informally, "routers-with-side-effect") into this 
architecture, and we still are not entirely satisfied with the resolution
presented in section \ref{sec:routers_with_side_effect}. Our frustration is
exacerbated by the fact that this L2-mangling does not meaningfully contribute
to the overall functionality implemented within a middlebox, but is simply an
artifact of legacy network construction and operation. We urge middlebox
designers to design middleboxes as bridges, not routers.

\paragraph{For SDN architects and switch designers.} We believe that the
degree to which the OpenFlow specification incorporates semantic understanding
of upper-layer protocols (such as IP and TCP) is excessive, and results in 
diminished flexibility and increased maintenance cost. Although we utilize
the MAC address header fields, the use to which we put them is decidedly not
for storing addresses; the semantic meaning attached to those fields by 
OpenFlow is effectively a legacy meaning rendered obsolete by SDN. However,
the restrictions of active switching (i.e. path length, port identifier size)
are all a direct result of that legacy meaning--- specifically, that the
addresses used in the 802.1d MAC are six bytes in length. We suggest that 
primitives matching and acting on bit-strings at given offsets into the packet
would result in a more flexible protocol, and that ease-of-use concerns could
be mitigated by incorporating the upper-layer protocol semantic awareness into
a controller or switch programming library.

We also note that register and rewrite actions vastly increase the 
flexibility of software-defined networking, and, thus, the availability and
feasibility of solutions to challenges within the networking space. We
encourage their broader availability.

\paragraph{For developers of SDN applications.} We note that there is a
striking similarity between the logic commonly programmed into SDN switches,
and the behavior of legacy ("flood-and-learn") switches. In many applications
that we have observed, it appears that the primary contribution of SDN is to
eliminate the need to flood and to allow forwarding decisions to be made on
upper-layer headers. However, as the specificity of match rules increases, a
greater number of such rules are required to describe policy for the same
volume of traffic, and the size of said match rules are significantly larger
than those of the rules learned by legacy switches.

As a result, we believe that this common approach to the construction of 
software-defined networks does not result in the promised efficiency gains of
SDN, and may, in fact, be less efficient than legacy networking.\footnote{At
least in terms of space.} We encourage the developers of SDN applications to
explore techniques that do not result in flow-table explosion on the order of
the number of flows.


\section{Conclusion}
This paper presents "active switching", a novel technique for the
construction of software-defined networks. Active switching is a general 
descriptor for any technique where flow-state is embedded into traffic, rather
than maintained in flow-table entries; and where switches' flow-tables 
act as transition functions modifying flow state.\footnote{There are 
intriguing theoretical implications to software-defined
networking, especially vis-a-vis register extensions. We think it intuitive,
for example, that such a network of sufficient size could be programmed so
as to recognize the class of regular languages. A full exploration of these
implications is left to future work.}
We have described the use of active switching to solve a number of issues, 
broadly characterized as "traffic-steering," across a variety of network 
topologies; although we suspect that the technique has far broader applications 
that are left for future work. We have shown that this technique can effect 
behavior from a software-defined network that would be challenging or 
impossible to implement based solely on "match-and-forward" logic, and that 
this technique can result in a dramatic efficiency gains.

{\footnotesize \bibliographystyle{acm}
\bibliography{citations}}

\end{document}